\documentclass[epsf,psfig,preprint2]{aastex}

\begin{document}

\title{LITHIUM DEPLETION IN PRE-MAINSEQUENCE SOLAR-LIKE STARS} 

\author{L. PIAU and S. TURCK-CHIEZE}
\email{piau@cea.fr, cturck@cea.fr}

\affil{CEA/DSM/DAPNIA/Service d'Astrophysique, CE Saclay,
 91191 Gif-sur-Yvette Cedex 01, France\\
}

\begin{abstract}
We examine the internal structure of solar-like stars in detail
between 0.8 and 1.4 $M_{\odot}$ and during pre-main
sequence phase. Recent opacity computations of OPAL along with a new 
hydrodynamical mixing process have been considered. We also introduce 
up-to-date nuclear reaction rates and explore the impact of accretion,
 mixing-length parameter, non-solar distributions among metals
and realistic rotation history. We compare models predictions of lithium depletion to the 
$^7Li$ content observations of the Sun and to 4
young clusters of different metallicities and ages. We show 
that we can distinguish 
two phases in lithium depletion: 1- a rapid nuclear destruction in the T-Tauri
phase before 20 Myrs : this is independent of the mass used within our range but largely dependent
on the extension and temperature of the convective zone, 
2- a second phase where the destruction is slow and moderate and which is largely dependent 
on the (magneto)hydrodynamic instability located at the base of the convective 
zone.

In terms of composition, we show the interest on considering  
helium and especially the mixture of heavy elements : carbon, oxygen, silicium
 and iron. We outline the importance of O/Fe ratio. 
We note a reasonable agreement on lithium depletion for the two best known
cases, the  Sun and the Hyades cluster for solar-like stars.
Other clusters suggest that processes which may partly inhibit the predicted premainsequence
depletion cannot be excluded, in particular for stars below $\sim 0.9M_{\odot}$. Finally we 
suggest different research areas such as initial stellar models and more realistic atmospheres
 which could contribute to a better understanding of this early phase of evolution
 and which should become the object of subsequent research.

\end{abstract}


\section{INTRODUCTION}
Dynamical effects have been mainly ignored in classical stellar evolution 
during several decades even if they have been explored theoretically 
(Zahn 1974, 1992). Nowadays, helioseismology provides observational constrains
on such effects and therefore allow us to
begin to introduce them in the description of main sequence stars (Frolich
 et al. 1997; Gabriel et al. 1997; Kosovichev et al. 1997). It permits observations of
 the solar convective layers complex motions. Moreover acoustic mode determination allows the
 extraction of the rotation and sound speed profiles down to the energy-generation core 
(Kosovichev et al. 1997; Dziembowski 1998; Turck-Chi\`eze et al. 1997). 
Meridional circulation begins to be accessible to the solar seismic observations and some 
(magneto) hydrodynamical instability has been put in evidence at the base of the convective 
zone. In order to see the consequences of such a process, we have focused our attention on 
two elements $^7Li$
and $^9Be$ which are destroyed in stars from the center to regions located slightly 
below the natural transition between transport of energy by radiation and by convection. 
Lithium surface abundances history is directly related to physical
 processes in this region which is at the same time 
important for understanding dynamos in stars and also probably the role of the 
internal magnetic field in general. Brun, Turck-Chi\`eze, \& Zahn (1999) have invoked the 
hydrodynamical instability proposed by Spielgel \& Zahn (1992) to explain both recent
helioseismic results and the solar lithium depletion. They introduced a 
turbulent term in the equation of diffusion. This term is directly related to local
 rotation and differential rotation. Evolution of such a mixing term with 
recorded surface rotation history of sun-like stars in open-clusters finally explains the present 
observed solar photospheric element abundances. It also produces the right order of 
magnitude of lithium destruction as a function of time for open-clusters older than a
 Gyr (NGC 752, M67).

In this paper we focus on the different ingredients of the
 mainsequence (MS) and above all premainsequence (pre-MS) structural 
evolution of solar-like stars through the excellent indicator which is the $^7Li$ surface 
abundance. From that viewpoint stellar evolution models have received a lot of attention for
 a long time. Almost forty years ago pre-MS evolution was already proposed to 
explain the low solar $^7Li$ abundances relative to the solar system value
 as well as field stars and Hyades depletion pattern with temperature 
(Bodenheimer 1965). Bodenheimer described the main features of lithium
 burning in pre-MS. Accurate $^7Li$ observational data for open clusters have similarly
been obtained for long (Zappala 1972). They have outlined the complex history of this 
element and suggested depletion with age on MS. In the past twenty years many theoretical
and observational works have addressed the topic.
These investigations suggested that the understanding of lithium abundance might be related to 
 many different non-standard processes extending
 from microscopic diffusion (Michaud 1986) to large 
scale mixing through rotation (Baglin et al. 1985) angular momentum evolution 
(Pinsonneault et al. 1989) or 
internal waves (Schatzmann 1993). During pre-MS lithium depletion strongly 
depends on typical temperatures within convection zones and therefore input physics (Profitt \&
Michaud 1989). Until today there is no safe explanation of lithium history among 
solar-like stars. It is very likely that various phenomena are indeed responsible
 for pre-MS and MS $^7Li$ evolution as well as G-type stars depletion pattern and lithium dip. 
This controversial subject therefore remains a broad and active area of research 
(D'Antona \& Mazzitelli 1994, 1997; Ventura et al., 1998; Palla 1999; D'Antona,
 Ventura \& Mazzitelli, 2000). 

This work intends to show results involving recent opacity and nuclear reaction
 rates determination. We consider possible variations of composition and non solar
 repartition among metals as in the next decade many improvements in the knowledge of 
the detailed photospheric composition of the young stars are anticipated.
 We envision variations of mixing-length parameter with age and a new hydrodynamical
 instability along with a realistic rotation evolution. We outline the impact of 
such improvements on the topic of early phases of evolution which
is meant to be better constrained by future seismic informations.
 In section 2, we describe the bases of the 
stellar models we use and examine the solar case in details. The role of the different elements 
through their composition and the opacity coefficients is discussed in details in section 
3. Section 4 is dedicated to convection \& accretion and section 5 to rotation. 
Finally we give some perspectives in section 6. 
 
\section{PRE-MAIN SEQUENCE EVOLUTION OF SOLAR-LIKE STARS}

\subsection{Physics and global evolution}

In this study, we consider models in the mass range from 0.8 to 1.4
 $M_{\odot}$ and different compositions corresponding to the Sun, the Pleiades, and the 
Hyades. Our starting point is a fully adiabatic polytrope with a central temperature
 of $\sim$ 3 $10^5$ K and a radius of 20 $R_{\odot}$ for $1M_{\odot}$ star. So 
the computation begins  prior to deuterium burning which corresponds to the observed 
birth-line (Stahler 1988). We observe that models are still fully convective
 polytropes when the star reaches the birth-line with a radius of about  6 $R_{\odot}$ so 
we expect no consequences of this early-phase computation
except the idea to take as an hypothesis an initial polytropic structure.

Models have been computed using the CESAM code (Morel 1997) and all the updated physics 
useful for the refined solar models (see Brun, Turck-Chi\`eze, \& Morel 1997). For the 
reaction rates of the pp chain and CNO cycle, we consider the compilation of Adelberger et al.
(1998) and for the $^7Li (p, \alpha) ^4He$ the work of Engstler et al. (1992). 
We use OPAL equation of state (Rogers, Swenson, \& Iglesias 1996) and 
opacities (Iglesias \& Rogers 1996) above 5800K. For lower 
temperatures the OPAL equation of state is replaced by 
the MHD equation of state (Mihalas, Dappen, \& Hummer 1988) and the opacities are from 
Alexander \& Ferguson (1994). Interpolation of opacities
 are performed using v9 birational spline package of Houdek (Houdek \& Rogl 1996).
 The atmosphere is connected to the envelope at optical depth 10 where 
the diffusion approximation for radiative transfer becomes valid (Morel et al. 1994).
 The opacity atmospheric is Rosseland mean opacity extracted from 
Alexander \& Ferguson (1994) or Kurucz (1992) sets. We therefore do not
use grey approximation. Surface 
boundary conditions are $\rho = 3.55\,10^{-9}g.cm^{-3}$ when $\tau = 10^{-4}$.
Standard mixing-length theory is applied and the convection zone is 
completely chemically homogeneous.

\begin{figure}
\centering
\rotatebox{90}{\includegraphics[width=6cm]{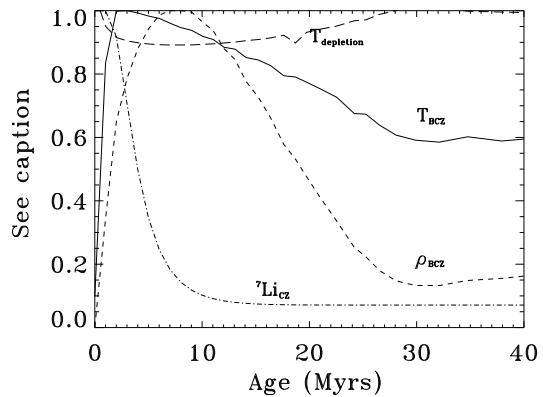}}
\caption{ 
Time evolution of the temperature (continuous line) and density (dashed line) 
at BCZ for a 1 ${M\odot}$ star of solar composition. Both temperature and density are 
normalized to their maximum values, respectively $3.94\times10^6$ K and $1.97 \rm g/cm^3$. 
We have also mentioned the density dependence of the temperature for a characteristic 
burning time of 1 Myrs (long-dashed line) and the lithium photospheric fraction 
(dot-dashed line) normalised to maximal value $9.3\times10^{-9}$ in mass fraction (ie 3.2 dex).
 }
\end{figure}
During the pre-MS, stars are in quasi-hydrostatic equilibrium
and slowly contract towards ZAMS within a time scale comparable to their Kelvin-Helmoltz
 time-scale. Defining ZAMS as the age where the thermonuclear hydrogen fusion provides 
99$\%$ 
of stellar energy the pre-MS lifetime varies from 30 Myrs for a star of 1.4 $M_{\odot}$ 
Pleiades composition  up to 100 Myrs for a 0.8 $M_{\odot}$ star of the Hyades composition. 
The Sun lies in between, with a pre-MS of $\sim$ 50 Myrs. The transition 
from fully convective object to radiative core structure depends on mass and composition. 
Higher mass stars contract faster, increase internal temperatures faster
 and so decrease radiative thermal gradients faster. Moreover lower metallicity 
accelerates contraction and
decreases opacities which also favor radiative stratification. Therefore lower 
metallicities and higher masses give rapid raise to radiative core. A massive (1.4 $M_{\odot}$) 
Pleiades composition star develops radiative core after 0.85 Myrs and
 a 0.8 $M_{\odot}$ Hyades composition star after 3.5 Myrs. In the solar
 case, radiative core exceeds 1$\%$ of total mass at 1.8 Myrs and
 then fastly accelerates in mass and radius to reach dimensions very
 close to its actual dimensions at 25 Myrs.

Due to the swift convective
 movements, surface matter is repeatedly exposed to physical conditions that prevail in deep
 interior. Figure 1 illustrates this point in showing the time dependent evolution of the 
thermodynamical quantities (temperature and density) at the base of the 
convection zone (hereafter
BCZ) in regard to the photospheric lithium for the case of a young Sun. At the beginning 
of the considered evolution BCZ (which coincides with the stellar center) is too cool to allow lithium 
burning. As the star 
evolves on pre-MS, deep regions of convection zone temporally exceeds $^7Li$
 burning point in typical stellar conditions ($\sim 2.5\,10^6K$).
$^7Li$ therefore offers a direct insight over stellar internal structure and
 evolution as it is extremely sensitive to the appearance of the radiative core.
 We note that early $^7Li$  
depletion occurs in the very beginning of pre-MS. Effectively, for 1$M_{\odot}$,
the BCZ temperature increases 
rapidly from less than  $10^6$ K to 
approximately 4 $10^6$ K at 2 Myrs, the density increases also (but not quite simultaneously) up 
to  2 g/$\rm cm^3$ between 7 and 8 Myrs. Then they slowly sink towards values which are 
very near from the conditions of the BCZ of the present Sun. This 
evolution results in $^7Li$ burning from ~2 to ~20 Myrs. 
The way depletion occurs presents several difficulties. 
First, depletion takes place just at 
the BCZ as convection rapidly recedes in 
mass-fraction passing from the whole star mass at 1 Myr to 
less than 20\% 20 Myrs later. Moreover depletion 
typical time evolves rapidly and decays to very low values compared to time 
scale evolution of the stellar structure down to $\sim 10^5$ 
years at BCZ. These pecularities impose small time step and very thin meshes.
\\
Computations suggest that one third difficulty must arise. 
Early radiation zone stratification hardly differs from convection
 stratification. In other words the newly stabilized
medium is (and stays for at least 10 Myrs) close to convective instability.
We can illustrate this using polytropes : the adiabatic polytrope
($P \alpha \rho^{1+1/n}$ with n=3/2) is 'harder' than its
radiative counterpart ($P \alpha \rho^{1+1/n}$ with n=3).
Radiation zone which represents half of the stellar mass 
at 10 Myrs, should there be more concentrated in the core. Figure 2
shows the contrary. The fact that radiation 
energy transport replaces convective motions does
not mean that radiative stratification immediately establishes.
At 10 or 20 Myrs the stratification in radiation zone
is still very close to adiabatic stratification. This has consequences
on rotation evolution as we shall see in section 5. We believe this 
property to be related to Kelvin-Helmoltz time value. The stellar interior needs 
at least this long to redistribute thermal energy and evolve from convective
to radiative structure. This suggests that little modifications in stellar structure or
new phenomena taken into account could easily change BCZ position at times from 10 to 20 Myrs.
 We may for instance suppose that even a low energy density magnetic 
field would stabilize or destabilize the region. This point leaves a priori room for
many non-standard mechanisms that will have to be investigated and that will probably lead to 
substantial variations in pre-MS surface $^7Li$ depletion (Ventura et al. 1998).

\begin{figure}
\centering
\rotatebox{90}{\includegraphics[width=6cm]{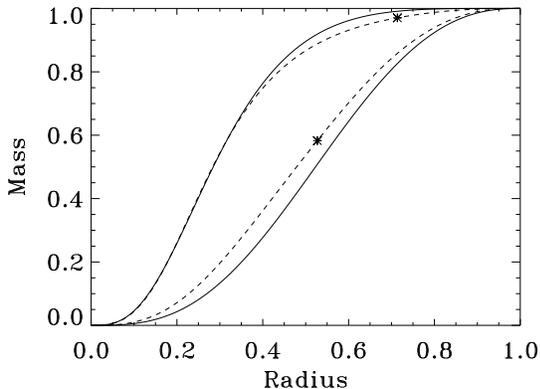}}
\caption{ Solid lines present the mass distribution of polytrope vs radius
for a 1$M_{\odot}$ star. Upper solid line corresponds to a polytrope 
of index n=3 (representative of radiative stratification), lower solid line
 corresponds to a polytrope of index n=3/2 (adiabatic stratification). Dashed-lines 
present model stratifications at 10 Myrs (lower line) and 30 Myrs
(upper line) for solar composition and mass. Crosses on 
these lines are BCZ. It is note worthy that at 10 Myrs stratification is still
 almost adiabatic albeit more than 50\% of stellar mass is in the radiation zone.}
\end{figure}

\begin{figure}
\centering
\rotatebox{90}{\includegraphics[width=6cm]{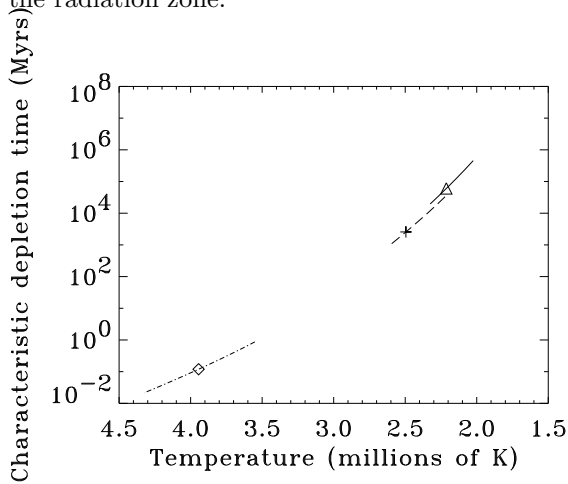}}
\caption{ Temperature dependence of the
$^7Li$ depletion characteristic time around BCZ
 for  a diffusive solar model (X=0.70821, Y=0.2722). 
The continuous line corresponds to the solar age, dashed line to
the ZAMS position, and dot-dashed line to a 3 Myrs star. Triangle,
 cross, and diamond represent the positions of the BCZs.
 }
\end{figure}

\subsection{ Adjustment of evolution parameters for $^7Li$ burning}

$^7Li$ reaction rate
 is extremely sensitive to temperature. 
 For typical pre-main sequence conditions, at the maximum temperature of BCZ
 (3.5 to 4 $10^6$ K) this rate varies as $T^{18}$ to $T^{19}$.
The depletion is therefore very sensitive to conditions 
near the base of the convection zone (figure 3). Consequently,
the photospheric $^7Li$ depletion strongly depends not only on  the mass of the convection 
zone but also on the precision of the mass shell division at BCZ. This point is essential 
to perform valid 
integrations. We have adjusted the mesh in order to limit the variation of the reaction 
rate at BCZ between two layers to 15 \%, this corresponds typically to meshes of 0.005 
$M_{\odot}$ or 0.003 $R_{\odot}$. Moreover, the typical 
$^7Li$ pre-MS burning requires also
a very 
 accurate resolution relative to time integration. 
Figure 3 shows the evolution of the characteristic time along the evolution. At 3-4 Myrs 
we find a characteristic destruction time at BCZ of only $8.10^4$ to 
$1.3\,10^5$ years (depending on composition). So we have adjusted the typical  
time step to be a factor 10 lower than the characteristic depletion time. In figure 3 we have shown three 
conditions corresponding to the maximum of the temperature at BCZ (3 Myrs), to the 
arrival on ZAMS and to the present age of the Sun. For
each condition, the depletion time varies by approximately one order of magnitude from BCZ 
to the right end
of the plot but this extension corresponds to only 10 $\%$ variation of the mass of the 
convection zone.

\begin{table}[ht]
  \begin{center}
    \caption{ Ratio of lithium fraction to initial fraction after pre-MS depletion 
phase within solar composition stars (see text
for details). Our results (PTC01) are compared to previous ones of D'antona \& Mazzitelli
 1984 (D'A\&M84), Profitt \& Michaud 1989 (P\&M89) and Ventura et al. 1998 (V98). 
P\&M89a and P\&M89b respectively corresponds to Z=0.0169 and 0.024. For $1.1 \& 
1 M_{\odot}$ no depletion is predicted on MS. For $0.9 M_{\odot}$ 
star D'antona \& Mazzitelli do not predict depletion on MS but Profitt \& Michaud 
1989 and us predict it. For these low mass stars lithium fraction therefore refers 
to a fixed age of 70 Myrs. Results of Ventura et al. 1998 are provided here only for 
solar mass stars as they do not give results for other masses in MLT framework.}\vspace{1em}
    \renewcommand{\arraystretch}{1.2}
    \begin{tabular}[h]{lccc}
      	Mass ($M_{\odot}$) & 0.9 & 1 & 1.1 \\
      \hline
	D'A\&M84 & 0.664 & 0.871 & 0.949 \\
      \hline
	P\&M89a & 0.467 & 0.679 & 0.823 \\
	P\&M89b & 0.118 & 0.317 & 0.535 \\
      \hline
	V98 & \ & 0.026 & \ \\
      \hline
	PTC01 & 0.002 & 0.071 & 0.244  \\
      \hline
      \end{tabular}
  \end{center}
\end{table}

\subsection{Solar lithium burning during
 pre-mainsequence}

\begin{figure}[ht]
\centering
\rotatebox{90}{\includegraphics[width=6cm]{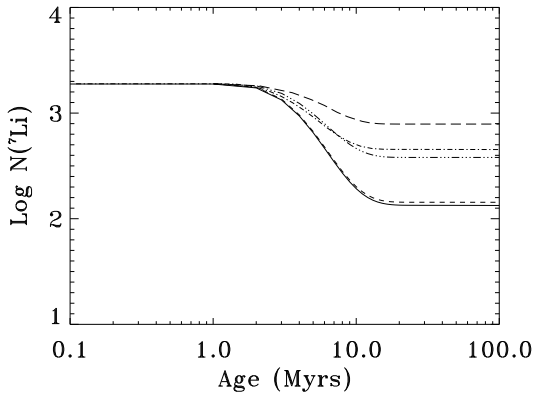}}
\caption{Time dependence of the photospheric $^7Li$ abundance for a 1$M_{\odot}$ star
 with solar initial composition for different computations: a diffusive calibrated model  
(Y=0.2722 and Z=0.01959) with a
time step adjusted to $^7Li$ BCZ burning time (continuous line), idem but the time step
imposed by the structure of the star (dot-dashed line), a nondiffusive noncalibrated model 
 with a time step adjusted to $^7Li$ BCZ burning time (dashed-line), idem but the time step
imposed by the structure of the star (long-dashed line). 
The calibrated  nondiffusive model with Y=0.2648 and Z=0.01763 with
correct time step is represented by (-...-...) line and shows the effect of the composition}
\end{figure}

Pre-MS $^7Li$ depletion within stars of roughly solar mass and composition  
has been estimated
in various previous investigations (Bodenheimer 1965; D'Antona \& Mazzitelli 1984 ;
 Profitt \& Michaud 1989; Ventura et al. 1998). These three last studies 
suggest lithium depletion to increase as input physics is updated. 
D'Antona \& Mazzitelli 1984 find that a solar mass 
star depletes $\sim13\%$ of its initial lithium. The authors choose Y=0.23, Z=0.02
and $\alpha_{mlt}=2$. Profitt \& Michaud 1989 find that a solar mass 
star depletes from $\sim32\%$ to $\sim69\%$ of its initial lithium. These 
values correspond to the same $\alpha_{mlt}=1.5$ but make
 different assumptions on the composition. The first
one refers to Y=0.28, Z=0.0169 as the second one refers to Y=0.28, Z=0.024.
Profitt \& Michaud suggest that difference to D'Antona \& Mazzitelli
depletion rates is probably due to opacities improvements. They moreover 
remark that hte gap would get even wider if they were to use 
D'Antona \& Mazzitelli parameters (Y and $\alpha_{mlt}$). 
The gap between 1984 and 1989 results increases towards low masses and 
 gets even wider if one considers very recent work from Ventura et al. (1998).
 A model from these authors having solar mass and composition (Y=0.28 and Z=0.02) 
and using MLT $\alpha_{mlt}=1.55$ brings initial lithium fraction 3.3 dex 
down to 1.72 dex (a decrease of a factor $\sim 40$). Altough a bit larger, 
such depletion is the same order of magnitude as what our computations suggest. 
The present work therefore confirms the tendency of depletion to increase
 with updated physics. We find that lithium lessens by roughly a factor ten
in solar mass and composition star (we take Y=0.2722, Z=0.01959, $\alpha_{mlt}$=1.766).
Opacities must partly be responsible for this evolution but new estimations
of nuclear reaction rates should also be responsible for this increase. The Engstler 
et al. (1992) $^7Li(p,\alpha)^4He$ astrophysical factor is indeed increased by 30 \% in 
comparison with older values (see BTZ99). Finally the correct attention to the adapted 
timestep to lithium depletion rate is a third possible source of dispersion (as shown below).
Table 1 compares our computations to previous ones.

Figure 4 presents the results for the evolution of the Sun in pre-mainsequence.
Following the work of Brun, Turck-Chi\`eze, \& Zahn (1999, hereafter BTCZ99)
we have computed 
models of the Sun, in introducing (or not introducing) microscopic diffusion. Depending on this 
choice, the initial composition Y= 0.2722, Z= 0.01959 (or Y= 0.2624, Z= 0.01763) is 
adjusted to get the correct luminosity and radius at the present age, calibrated with an 
accuracy better than 4$\,10^{-4}$. Figure 4 shows that the microscopic 
diffusion has no effect on the pre-mainsequence. It is too slow a 
 process as it results in a change of photospheric helium and metals 
abundances of only 10\% along the  whole life of the Sun (Turcotte et al. 1998).
We have to note here a crucial point in $^7Li$ pre-MS depletion. Sun-like 
stars change much faster on pre-MS than on MS because evolutionary time
scale is primarily related to contraction (hence to Kelvin-Helmoltz time
$\tau_{KH}$) which is much smaller than nuclear reaction time ($\tau_{N}$).
Yet one should not expect any calculation on lithium during early pre-MS 
to provide correct results without some cautions. Despite $\tau_{KH}<<\tau_{N}$
 forces to adopt in the code small temporal steps these are much larger than
$^7Li$ burning time at BCZ ($\tau_{^7Li}$) at these ages. The following 
comparisons illustrates this.
In diffusive models $^7Li$ depletion is increased by a factor of 3.5 
between a star where time evolution is always less than one tenth of $\tau_{^7Li}$
 and a model where the time step is given by the evolution of the structure. 
For nondiffusive models this ratio goes up to 5.3. Once time step is well below $\tau_{^7Li}$
no change in $^7Li$ depletion is seen at a
 given composition for various timesteps and with or without
diffusion. In every following computations we take time step 
to be $\frac{\tau_{^7Li}}{10}$ as long as lithium is rapidily depleted (prior to $\sim 30$ Myrs).
Furthermore burning rate is mass-averaged over convection zone.
We have checked that the results are robust to the choice of time step. If we decrease time step 
down to $\frac{\tau_{^7Li}}{30}$ : $^7Li$ depletion increases of only 3.5 \%.
With time step adjusted for a good treatment of $^7Li$ burning, the difference of 
about 0.5 dex after 20 Myrs (a factor 3 in destruction) between diffusive and nondiffusive
 models is only due to the change of initial composition. 
This shows the great dependence of the lithium burning
on the composition (see next section). Such precise computation of 
the lithium burning in the pre-mainsequence was not included in the previous work of 
BTCZ99.

Of course the notion of calibration  between diffusive and nondiffusive models
is not justified in the study of young clusters but the great sensitivity of this 
calibration shows already the difficulty of predicting
 lithium burning in pre-mainsequence stage.

$^7Li$ is generally not the only probe of the stellar internal structure but
$^6Li$ is depleted at lower temperature and is of no help here. We have observed that $^9Be$ 
burning gives similarly poor indications. In fact, computed stellar models of 0.8  
to 1.4 $M_{\odot}$ show a $^9Be$ depletion of less than 0.04 dex whatever the cluster
 membership (from Pleiades to Hyades). 
 
\section{THE ROLE OF THE DETAILED COMPOSITION}

\subsection{Open cluster evolution}

Open clusters allow a direct test of $^7Li$ evolution with time and/or 
composition. In this paper we focus our attention on two young and two 
midle-age clusters : Pleiades and Blanco I ($\zeta$ Sculptoris) for the former 
Hyades and Coma Berenices for the latter. The objective is to distinguish 
metallicity from age effects. Blanco I and the Hyades apparently are metal-rich clusters 
with respectively $[Fe/H]=0.127\pm0.022$ (Boesgaard \& Friel 1990) and 
$[Fe/H]=0.14\pm0.01$ (Jeffries \& James 1999) however Blanco I is much younger than the 
Hyades and with an age around 50 to 90 Myrs comparable to the Pleiades cluster 
which is estimated to be 70 Myrs old (Patenaude et al. 1978). 
On the other hand Pleiades and Coma Berenices clusters have metallicities 
very slightly below solar one with $[Fe/H]=-0.034\pm0.024$ (Boesgaard \& Friel 1990) 
and $[Fe/H]=-0.052\pm0.026$ (Friel \& 
Boesgaard 1992) respectively but Coma has an age of 500 Myrs comparable 
in age to the old Hyades cluster of $\sim 600 Myrs$ (Perryman et al. 1998). 
It should be mentioned that composition data are undoubtedly more reliable 
in the case of Hyades or Pleiades than in the case of Blanco I which is
more distant ($\sim 250\pm30 pc$) than all other clusters and has been less studied.

Unlike the sun, microscopic diffusion has a negligible effect in the evolution
 of these young or midle-age clusters. We compute a decrease of 1.3 and 0.8 
\% for helium and metals in Hyades solar mass star case. As such variations always 
remain largely smaller than error bars over metallicity or helium content it 
is immaterial to study initial composition effect resulting from diffusion.
This is particularly true for lithium. Let us however remark that microscopic 
diffusion time appears roughly ten times shorter for F type stars. We compute a
 decrease of respectively $10\%$ and $6\%$  for helium and metals 
in $1.4M_{\odot}$ ($T_{eff}\sim 6600 K$) at Hyades age. However 
at an age of 50 Myrs decrease of helium and metals are only $0.6\%$ and $0.4\%$
 so that it is not plausible that microscopic diffusion changes early lithium 
history even in case of slightly more massive stars than the sun.

In this study, we do not consider different initial $^7Li$ 
abundances. In all four clusters stars more massive than 1.4 solar masses 
(Teff$>$6900 K) exhibit the same $^7Li$ content of 3.2 to 3.3 dex
 for Pleiades, Hyades, (Soderblom et al. 1993) and Coma (Boesgaard 1987) . Blanco I 
depletion pattern is furthermore identical to Pleiades' one (Jeffries 
1999). The early F-stars abundances remain unchanged and are
compatible both with very young T Tauri stars (Magazzu, Rebolo, \& Pavlenko 1992) 
and the interstellar medium present $^7Li$ value (Knauth et al. 2000). Indeed,
 it seems that there was no significant evolution in galactic gas 
$^7Li$/hydrogen ratio over the last 1.7 Gyr (Hobbs et al. 1988) so that the 
initial $^7Li$ does not vary from Hyades formation time until today. For all the 
clusters we take the same standard value of 3.27 dex for initial $^7Li$ abundances.

In the following we will  investigate the effects of the composition on the lithium 
burning in separating the effects of deuterium,
 helium, and metals. 
\\
\\
\subsection{Sensitivity of the lithium burning to the deuterium composition}
 Deuterium, the most fragile element, is depleted around $5\,10^5$ K in stellar interiors.
The galactical evolution leads to a continuously decrease  
because of astration. Recent measurements show an abundance of 
$(D/H)_{ISM}=1.46\pm0.09\times10^{-5}$ (Piskunov et al. 1997) or 
$(D/H)_{ISM}=1.60\pm0.09\times10^{-5}$ in direction of Capella (Linsky et al. 1995) whereas
presolar value is estimated as $(D/H)_{pre\odot}=3.01\pm0.17\times10^{-5}$
by Gautier \& Morel (1997). Young clusters initial abundances are probably
 the same as present ISM one.

During contraction, the stars experience several stages 
of light element burning. First, deuterium burning (T = $5\,10^5$ K)
significantly participates to the energy production. Palla \& Stahler (1991) have shown 
that for 1 $M_{\odot}$ star, deuterium burning stops the contraction at a radius of
$\sim 5-6 R_{\odot}$ which defines the so called birth-line (Stahler 1988). 
 The deuterium burning stops 
before 1 Myr when first phase lithium depletion starts. Consequently the star should have enough 
time to 'forget' its previous history. We checked this briefly in 
the framework of our models. For solar mass stars,
 deuterium begins to burn at the age $\sim 4\,10^{4}$ years during $\sim 2\,10^{5}$ years 
duration. Then lithium depletion starts after $1.4\,10^{6}$ years shortly
prior to radiative core appearence. We have considered three deuterium
mass fraction: the solar value, the actual ISM value and also a null value.
Such variations hardly change lithium photospheric abundances evolution. Models 
having Hyades or Pleiades composition, show
maximum variations between them of $\sim 10\%$ (0.04 dex)
 up to clusters or solar ages. When deuterium fraction decreases, 
lithium depletion slightly increases. 

\subsection{Sensitivity of lithium burning to the helium content}
$^7Li$ evolution is sensitive to metals but also to helium mass fraction. 
The solar photospheric helium has been determined by helioseismology to
be 0.246-0.249 (Basu \& Antia 95). 
Including microscopic diffusion (partly inhibited by turbulence) in calibrated solar models provides a 
mean to reach the initial solar helium content, estimated to be 0.2722 (BTCZ99). For open 
clusters there is no recent 
helium abundance determinations for any  studied cluster except the 
Hyades. A ZAMS low mass stars fit of Hyades suggests the cluster helium 
fraction to be $0.26\pm0.02$ (Perryman et al. 98). This value is confirmed by the 
position of a number of Hyades binaries in M-L diagram (Lebreton 2000). We 
adopt this value in our calculations. However other calculations (Pinsonneault et al. 
1998) suggest an higher value of 0.283. For the 
other clusters we use the usual helium 
to metallicity scaling law : $\Delta$Y/$\Delta$Z=$3\pm2$
as deduced from calibration of nearby visual binary systems (Fernandes et al. 1998).
In the same manner evaluations based on HII regions suggest an increase 
of helium content with metals. If we refer to figure 12 from Meyer 1989 we 
find that $\Delta$Y/$\Delta$Z lies between 1.5 and 6. Such a trend is 
not totally satisfactory as the Hyades do not seem 
to follow this law. In fact the use
of such a law requires a knowledge of both helium and 
metallicity in at least one site which is generally 
taken to be our solar system neighborhood.
But the solar system seems not to be representative
of 4.6 Gyrs ago mean trends of the ISM (Gies \& Lambert 1992) so it is possible that
deduced helium fractions are biased. Other calibrations
of the metal/helium relation coming from globular clusters and galactic
bulge measurements do not lead to similar helium fractions (Deliyannis, Demarque \& 
Kawaler 1990). Changes here are not so 
important, so we retain here solar system abundances as reference point. 
 
We have estimated the impact of helium variation 
on lithium depletion around solar composition, for Pleiades age, together with
 reasonable variations of the mixing-length parameter $\alpha$, and metal fraction.
The effect of the mixing length parameter is small, an increase of 5\% leads to a decrease 
of $^7Li$ content
by 20\%. On the contrary the impact of the composition is high: an increase of 
$\Delta$ Y= 0.025, leads to a decrease of lithium burning by 64\%, the corresponding 
$\Delta$ Z = 0.025/3 leads to an increase of lithium burning by a factor 300 and the 
correlated variation of helium and metallicity leads to an increase of lithium burning
by a factor 60. We address the metallicity dependence more precisely in the next chapter.
The anti-correlation between lithium depletion and helium 
is reported in previous calculations. D'Antona \& Mazzitelli (1984) analyse helium mass 
fraction 0.23 and 0.28. There the rate of lithium fraction remaining from their initial 
fraction to this helium mass variation ($\frac{^7Li}{^7Li_{0}}$/$\Delta Y$) 
is roughly 3.3 and 1.3 for $0.9$ and $1M_{\odot}$ stars respectively. Considering 
a helium variation ($\sim 0.05$) from 0.2624 to 0.32 we correspondingly find
rates 2.7 and 4.7 for $0.9$ and $1M_{\odot}$ stars. This is the same order of
magnitude as D'Antona \& Mazzitelli one although our dependence seems a bit larger.
The sensitiveness on helium fraction can easily be understood since opacities in stellar interiors 
reduce with helium content : less electrons are available for the same 
amount of matter. If we assume that helium content is a free parameter
we can speculate as to what level it should be increased to agree with
 the observations in
 Pleiades case. Lithium abundances being scattered for any given effective temperature
one has first to derive a mean value. Figure 5 (dashed line) shows a third degree polynomial
least square fit of observed abundances. We obtain helium fraction of
 0.36 for 0.85$M_{\odot}$ star, 0.32 for 1$M_{\odot}$, and 0.3 for 1.4$M_{\odot}$.
 These very high values seem difficult to justify. Recent work by 
Deharveng et al.(2000) on helium abundances in HII galactic regions shows
$He/H$ to be below 0.105 where it can safely be determined from $He^+/H^+$ ratios.
This  corresponds to helium mass fraction below 0.3. Another work 
 indicates helium mean mass fraction should be $0.28\pm0.02$ in the
galactic bulge (Minniti 1995) which is probably an upper limit to galactic
abundances as most chemically evolved stars populations are expected 
to be near the center of our galaxy. A second difficulty is that
 helium mass fraction would have to vary with mass by about 20\%. This
 certainly is not what we observe in Pleiades.
 Nevertheless, it is clear from this analysis that a proper determination of helium 
 is a crucial knowledge to understand the young clusters lithium evolution.

\subsection{Sensitivity of lithium burning to the metallicity}
We have computed several models 
to describe the 1 $M_{\odot}$ Hyades and
 Pleiades stars. These models rely on physics described in section 2.1.  
 Table 2 summarizes these results. We have calculated three 
types of models: the first one (A) uses metallicity deduced from the observation of iron
given in section 3.1
and $\alpha$ parameter calibrated on solar models (1.766). The second model (B) adopts
the same composition but three different values of 
$\alpha$ were used  to account for recent 2D hydrodynamical 
evaluations  (see section 4.1). The third model (C) is an 
extremal model in the sense that it adopts
 the lowest metallicity and highest helium content within
error bars in an attempt to cancel discrepancies between calculations
and observations. In the Hyades case we have added a supplementary 
model (D) introducing the value of 0.283 in helium mass fraction 
claimed by Pinsonneault et al. 1998. Other Hyades models have been computed, we
discuss them in more detail in section 3.5

\begin{table*}[t]
\caption{1$M_{\odot}$ stellar models for Sun, Pleiades, and Hyades compositions.
All models include microscopic diffusion. Several $\alpha$ parameter values
 have been considered : 1.766 is value assumed at solar age it is deduced from solar calibration.
$\alpha$=1.748 results from slightly different calibration when tachocline mixing
is taken into account.
 1.935 and 1.850 are values induced from hydrodynamical simulations (Ludwig, Freytag, 
\&Steffen 1999). For Pleiades composition
we take $\alpha$=1.935 before 15 Myrs and $\alpha$=1.85 from 15 to 22 Myrs. For
Hyades composition we take $\alpha$=1.935 before 20 Myrs and $\alpha$=1.85 
from 20 to 28 Myrs. Lithium dex fraction is
 presented at appropriate age. The last column provides effective temperature
to outline composition impact on it.}\vspace{1em}
\begin{center}
\begin{tabular}{lccccccc}\hline
 & Z &    Y  & $\alpha$ & $^7Li$ content & [O/Fe] & Tachocline & $T_{eff}$\\
 & (mass fraction) & (mass fraction) & & (dex) & & & (K)\\
\hline
Solar & & & & & & &	 \\
models & & & & & & &	 \\
\hline
Reference		& 1.959 $10^{-2}$ & 0.2722          & 1.766 &  2.0 & 0 & No & 5776 K \\
\hline
Tachocline 1		& 1.903 $10^{-2}$    &  0.2695   &  1.748 & 0.35 & 0 & Yes $\tau_{disk}=0.5$ & 5777 K \\
\hline
Tachocline 2		& 1.903 $10^{-2}$    &  0.2695   &  1.748 & 1.1 & 0 & Yes $\tau_{disk}=10$ & 5777 K \\
\hline
\hline
Pleiades & & & & & & &	 \\
models & & & & & & &	 \\
\hline
A			& 1.632 $10^{-2}$ 	& 0.2624          & 1.766 &   2.6  & 0 & No & 5717 K \\
\hline
B			  &  1.632 $10^{-2}$    &  0.2624   & 1.935, 1.850, 1.766 & 2.4  & 0 & No & 5715 K \\
\hline
C  		   &        1.535 $10^{-2}$ & 0.2679 & 1.766    &  2.75  & 0 & No & 5801 K\\
\hline
\hline
Hyades & & & &  & & &	 \\
models & & & &  & & &	 \\
\hline
A			& 2.367 $10^{-2}$ & 0.2633  & 1.766  &  0.9   & 0 &  No & 5450 K\\
\hline
B                       &  2.367 $10^{-2}$  & 0.2633 & 1.935, 1.850, 1.766 & 0.5 & 0 &  No & 5466 K\\
\hline
C			& 2.180 $10^{-2}$  &  0.28 & 1.766 &  1.75 & 0 & No & 5646 K\\
\hline
D 			& 2.28 $10^{-2}$  & 0.283 &  1.766 & 1.55 & 0 & No & 5632 K\\
\hline
E     &  1.57 $10^{-2}$  &  0.26  &  1.766 &  2.64   & -0.2 & No & 5802 K\\
\hline
F     &  1.94 $10^{-2}$  &  0.26  &  1.766 &  1.98  & -0.2 & No & 5638 K\\
\hline
G     &  1.57 $10^{-2}$  &  0.26  &  1.766 & 2.3    & -0.2 & Yes $\tau_{disk}=10$ & 5805 K\\
\hline
H     &  1.57 $10^{-2}$  &  0.26  &  1.766 & 1.75   & -0.2 & Yes $\tau_{disk}=0.5$ & 5804 K\\
\hline
\end{tabular}
\end{center}
\end{table*}

On the main sequence effective stellar temperatures evolve slowly. 
The temperature of an open cluster solar mass star will therefore 
be reliable independently on age uncertainties. Moreover these
temperatures do not vary much within composition uncertainties
 or $\alpha$ parameter different evaluation along early pre-main sequence
(see table 2). For instance the 1$M_{\odot}$  Hyades
 models exhibits a variation of less than 10 K around 
5450 K between 550 and 700 Myrs (idem for the  model including $\alpha$ parameter
effects on early pre-MS variation  but around 5500 K).
\\
We note in figures 5 and 6 that $^7Li$ depletion is too strong on pre-main 
sequence for solar-type stars with Pleiades or Hyades composition. This agrees with 
recent results from Morel et al. (2000) of the B component of $\iota\,$ Pegasi binary 
system. With an estimated age of 56 Myrs for the system and [Li]=2.69 dex for
the B component this 0.819 star is clearly underdepleted. 
This problem is pointed out in other recent studies. 
Ventura et al. (1998) using the full spectrum of turbulence of Canuto, Goldman, 
\& Mazzitelli (1996) prescription 
for modelling the convection  found also
a too strong depletion for solar composition stars, and similarly a very strong 
dependence on metallicity and mass.

 It is well known that open cluster stars exihibit an anti-correlation between 
effective temperature and lithium abundances. Moreover the dispersion 
in lithium abundances grows when temperature declines. This dispersion is too large to be 
due to abundance uncertainties (Soderblom et al. 1993). This is also too
large to be due to color errors as the vector of temperature 
errors is nearly parallel to the mean lithium-$T_{eff}$ trend in cool open 
cluster stars. Moreover this dispersion is already observed in 
$^7Li$ equivalent width as a function of color (Soderblom et al. 1993, Thorburn et al. 1993).
At a given effective temperature there is undoudtedly real star to star differences. 

Observations for each star give an effective temperature and a lithium abundance but
these quantities are not directly measured as they are deduced from photometric
 measurements such as (B-V) and equivalent width of a particular spectral
 absorption line. Measurements are always slightly scattered and moreover
there are unavoidable errors when doing conversions. For Pleiades
 open-cluster, Soderblom et al. (1993) find
 uncertainty for effective temperature of $\sim\,130 K$ and uncertainty for
$^7Li\,\sim 0.05$ dex. For the Blanco I cluster Jeffries \& James (1999)
give an effective temperature uncertainty of $\sim\,250 K$ and an abundance
 uncertainty of 0.11 dex. Figures 5 and 6 clearly show that such uncertainties 
are not large enough to recover agreement to computations. As we mainly 
analyze $1M{\odot}$ evolution, we concentrate on corresponding 
error-bars. It is generally found in litterature that these
stars verify $0.6<(B-V)<0.75$. In fact, lithium abundances are such steep functions
 of effective temperature that if
we use this criterium, we gather in the same sample stars with 
very different $^7Li$ fractions. Let us consider Hyades case. 
In Thorburn et al. (1993) data, the mean value of lithium content for  stars exihibiting 
$0.6<(B-V)<0.75$ is $N(^7Li)=2.0\pm0.6$ dex, but if we limit to
$5400 K<T_{eff}<5500 K$ which is what we get in computation, we get 
$N(^7Li)=1.3\pm0.1$ in the same data set. The small
 sample presently discussed greatly reduces 
differences between observations and numerical predictions.
Yet differences are not definitively cancelled out, and it is obvious from figure
5 and 6 that such possible misleading associations between stellar mass and 
(B-V) will not succeed in explaining the very large observation/theory gap concerning 
lower mass stars. 
\\
Table 2 shows that expected dispersions among metallicity or
 uncertainties on helium fraction can be responsible for dispersion
of lithium abundances. In Pleiades case,
the metallicity variation from
-0.034 dex (Z= 1.632 $10^{-2}$)
to -0.058 dex (Z= 1.535 $10^{-2}$) minimal value corresponds
 to 6$\%$ in metal mass 
fraction and produces 0.35 dex variation in photospheric lithium, which
is quite the order of observed dispersion for solar mass stars within this
cluster. In Hyades case, helium uncertainties
 at a level of $\sim 7\,\%$ correspondingly produce 0.6 dex $^7Li$ 
dispersion between models types (A) and (D). Composition variations are
therefore 
able to
explain $^7Li$ dispersion through pre-MS initial depletion
effect but mass accretion could contribute to it also (section 4).
If such variations explain lithium dispersions in young Pleiades or Coma
 clusters they also have to deal with middle-aged cluster observations. In Hyades
 dispersion in the $T_{eff}-^7Li$ relation has considerably decreased.

\begin{figure}
\centering
\rotatebox{90}{\includegraphics[width=6cm]{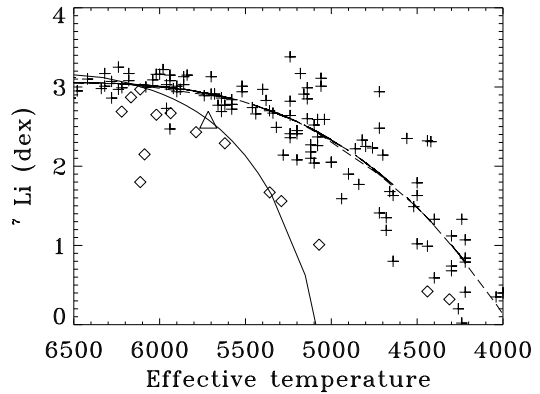}}
\caption{ 
Crosses : $^7Li$ abundances for members of the Pleiades (Soderblom 93 data). 
Diamonds : Coma Berenices (Jeffries \& James 99 data). Solid line is prediction of
 standard (corresponding to model Pleiades A) model of Pleiades composition. Small
 triangle on this line is 1$M_{\odot}$ star of pleiades composition. Dashed-line 
is lithium fraction interpolation in Pleiades.
 }
\end{figure}

\begin{figure}
\centering
\rotatebox{90}{\includegraphics[width=6cm]{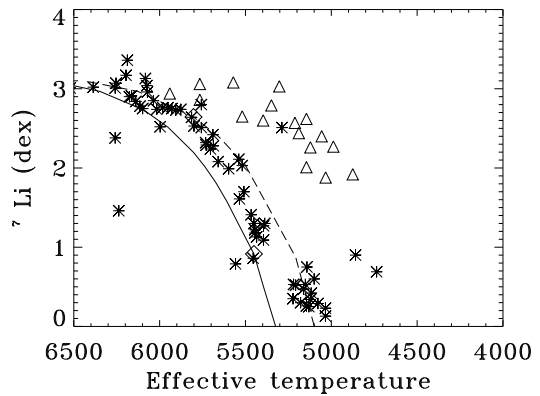}}
\caption{ 
Crosses : $^7Li$ abundances for members of the Hyades (Thorburn 93 data). 
Triangles : Blanco I (Jeffries data). Lines show model computation for different
 masses at $\sim 625$ Myrs. Solid line is prediction of standard (corresponding to
 model Hyades A) model of Hyades composition. Dashed line is prediction of case E
 model (see section on opacities). Small diamonds on both lines represent
1$M_{\odot}$ star.
 }
\end{figure}

\subsection{Opacity role on $^7Li$ burning}

Because of strong dependence of lithium pre-MS depletion on metallicity 
opacities effects in stellar interiors have been studied 
(Turck-Chi\`eze et al., 1993; Turck-Chi\`eze 1998;
Turcotte \& Christensen-Dalsgaard 1998). Solar-like star metal
opacities increase is responsible for
 the transition between radiative and convective energy transport. In the present Sun,
 the main contributors to the opacity at BCZ are oxygen and iron. They 
correspond respectively to 36 and 20 \% of the total opacity (table 3).
Situation is somewhat similar regarding pre-main sequence but the
principal metal contributors to opacity are not the same because the
 medium is denser and hotter. The roles of oxygen and iron
are of the same order of 20 \%, then neon, silicium, and magnesium each
represent approximately 10 \% of total opacity. 
\\
Up to now we have deduced metallicity from [Fe/H], in keeping 
the solar distribution inside the metals.
 There have been hints that solar abundances may
deviate from mean ISM abundances and that solar system would not be representative
 of general trends (Gies \& Lambert 1991). Regarding open clusters, similar situation
could appear even if direct determinations have to be considered cautiously due to
 chromospheric activity see Cayrel et al. (1985) for Hyades. Let us consider Hyades cluster.
 It is the closest and probably the best documented. 
Moreover it appears metal-rich. Hence Hyades 
stars experience amplified early pre-MS $^7Li$ depletion.
As we have already said, predominant metals in opacity generation are
iron, oxygen, neon, silicium, and magnesium. The full set of these 
metals represents more than $70\%$ of total opacity whereas any other accounted
 for metal do not contribute over $5\%$. In table 3 we give opacity 
contributions of 19 metals at BCZ for a 1$M_{\odot}$ star extracted from
 monochromatic calculations of Iglesias \& Rogers (1996). 
Those results are obtained the following way :
for every component we compute total opacity for the stellar plasma 
mixing excluding that component. Difference of opacity with the plasma
 including every components then allows to estimate to what level that
 element is responsible for opacity.
\begin{table}[ht]
  \begin{center}
    \caption{ Repartition of elements in opacity calculations at BCZ for solar composition in typical
	      main sequence ($\rm T\sim2.10^6K \rho\sim0.2g.cm^{-3}$) and pre-main sequence 
($\rm T\sim4.10^6K
		 \rho\sim2g.cm^{-3}$) conditions.}\vspace{1em}
    \renewcommand{\arraystretch}{1.2}
    \begin{tabular}[h]{lcc}
      	Element & MS & pre-MS \\
      \hline
	H & $20\%$ & $13\%$\\
      \hline
	He & $11\%$ & $6\%$\\
      \hline
	C & $8\%$ & $2\%$\\
      \hline
	N & $5\%$ & $1\%$\\
      \hline
	O & $36\%$ & $19\%$\\
      \hline
	Ne & $9\%$ & $13\%$ \\
      \hline
	Mg & $3\%$ & $11\%$\\
      \hline
	Al & $<1\%$ & $1\%$ \\
      \hline
	Si & $3\%$ & $12\%$ \\
      \hline
	S & $4\%$ & $5\%$ \\
      \hline
	Ar & $1.5\%$ & $<1\%$ \\
      \hline
	Ca & $2\%$ & $<1\%$ \\
      \hline
	Fe & $16\%$ & $22\%$ \\
      \hline
	Ni & $1\%$ & $2\%$ \\
      \hline
	For every following metal \\
	Na,P,Cl,K, \\
	Ti,Cr and Mn & $<1\%$ & $<1\%$ \\
      \hline
      \end{tabular}
  \end{center}
\end{table}

We now try to determine the detailed composition of the Hyades for relevant metals.
Edvardsson et al.(1993) find a tight correlation between [O/Fe]
and [Fe/H] in galactic disk stars,
\\
$[O/Fe]=(-0.36\pm0.02)\times[Fe/H]-(0.044\pm0.010)$
\\
\\
This relation once again suggests our Sun could 
be overabundant in oxygen when compared to other stars.
Moreover Garcia Lopez et al. (1993) claim oxygen to hydrogen metallicity
is close to $[O/H]=-0.07\pm0.05$ dex in solar-like Hyades stars. 
This result is qualitatively consistent with the previous relation.
 If we apply this law, Hyades appear now metal-poor in oxygen 
compared to the Sun, this is in agreement
with helium determinations from Perryman et al. (1998). Neon is not
detectable in solar-like stars photosphere but its ratio to oxygen measured in
HII regions is remarkably constant for different [O/H] (Meyer 1989). 
Indeed neon, silicium, and magnesium are expected to vary like oxygen because of their
common origin in SNII. Consequently we decide to take the same variation for 
silicium and magnesium than for oxygen. Let us just signal one caveat 
regarding silicium : Cayrel, Cayrel\,de\,Strobel, \&Campbell 1985 find $[Si/H]=0.16\pm0.05$. 
Carbon cannot be ignored at least on MS but its origins are
 still a matter of debate (Gustafsson et al. 1999). We let it vary like oxygen.

We have generated opacity tables for this non-solar metal distributions using
the devoted Lawrence Livermore National Laboratory website. By doing so we 
have chosen to model two cases. In the first one 
(case E), every metal varies like oxygen except iron-peak
elements; in the second case,  we let every metal vary like iron except the oxygen,
to test the  dependence of lithium depletion on oxygen (case F).
 In case E, which is the most realistic, lithium abundance
is 2.6 dex  (instead 0.9 dex) for a 1$M_{\odot}$ star.
 The shift of oxygen abundance from 0.127 to -0.07 dex 
 increases lithium fraction by a large amount of 1.1 dex. 
Then the shifting of every metal but iron-peak ones 
to the level of oxygen increases again lithium fraction 
by 0.6 dex. This dramatic increase in lithium
fraction is shown on figures 6 and 7. Lithium in solar-like Hyades stars becomes quite 
compatible with observations. There are two reasons for this.
First, the decrease in metal fraction reduces BCZ 
temperatures and secondly it increases effective temperature for any given mass.
Solar mass stars are shifted from $\sim 5450$ to $\sim 5800$ Kelvins. 
The large effect of metal might at first seem worrying. However one as to keep in mind
two points. Firstly metals determine roughly 80 \% of opacities at BCZ during pre-MS. 
Secondly lithium depletion is a very steep function of temperature (see figure 3). 
Furthermore we remark that such a large impact of metallicity on 
depletion is reported in other recent
studies (Chaboyer, Demarque, \& Pinsonneault 1995). Ventura et al. 1998 
use OPAL opacity table although a slightly older version. They
claim a decrease of 15\% in the value of Z (around Z=0.02) makes the $^7Li$
 abundance to vary by
almost two orders of magnitude. The present effect of oxygen (0.2 dex represents 
$\sim 60\%$) which is half of metal fraction is quite compatible with this last result. 

We note that even lower mass stars appear in good agreement with observations. 
Figure 7 outlines the importance of metals in lithium depletion. It also
shows that oxygen plays a more determinant role than iron for pre-mainsequence evolution.

\begin{figure}[ht]
\centering
\rotatebox{90}{\includegraphics[width=6cm]{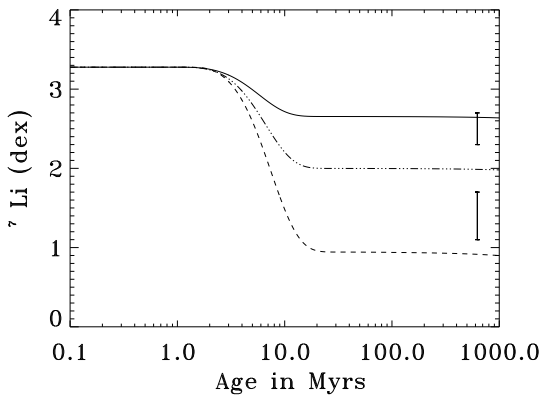}}
\caption{Predictions compared to observations for solar-like stars of Hyades composition.
 Solid line corresponds to Hyades' composition with every metal scaled on oxygen [O/H]=-0.07 dex
except iron-peak elements (case E). Dashed line corresponds to model where
all metal are scaled on iron [Fe/H]=0.127 dex (case A). For dashed 
dot line all metal are scaled on iron except oxygen (case F). 
Error-bars photospheric observed lithium are deduced from
Thorburn et al. (1993) data for solar-type stars considering  
$T_{eff}$ within 100 K around 5450 K 
(case A) and around 5800 K (case E). }
\end{figure}

If opacities are crucial at the base of convection zone they are also of first importance
in the atmosphere, as they determine depth where material becomes 
convective and this affects significantly the convection zone extension. 
The deeper convection starts in atmosphere the more efficient it 
is and the deeper it goes in the envelope. For low effective temperatures 
encountered on the Hayashi track many contributors to opacity have to be considered.
The lowest temperature we reach in computation is 3200 K and below 5000 K molecules are
 not negligible anymore. It is therefore likely that improvements of present models
 may be made when using new atmosphere models (Hauschildt, Allard, \& Baron 1999). 
Here we have mainly use Alexander \& Ferguson opacity tables (Alexander \& Ferguson 94)
and checked that there was no important discrepancy induced when using low-temperatures
 Kurucz opacity tables (Kurucz 1992). For case A composition star, final (and maximal)
 difference between otherwise similar models is less than 0.1 dex which is very small if
we recall global depletion factor is more than 2 dex in this case.  

In typical pre-MS conditions $H^-$ remains the main opacity source. The metals 
provide a larger part
of free electrons when temperature lowers. Alexander \& Ferguson opacity table
have been computed for solar composition and is not adapted for case E.
 We can evaluate average ionisation states
of all components in atmospheric thermodynamic conditions. For the coldest ($\sim3164 K$) 
and most diffuse ($1.5\,10^{-10} \rm g.cm^{-3}$) conditions, iron-peak elements
 and oxygen are respectively responsible for $\sim10$ and 0.25 \% of electron density.
When opacity evaluation changes from Alexander to OPAL table, temperature is logT=3.75
and density is typically a few 10$ ^{-7}$ g.cm$^{-3}$. We estimate iron-peak elements
 to be responsible for $\sim30$ \% of electron density and contribution of oxygen
being negligible.

In case E metal fraction is reasonably scaled by oxygen which is the main mass 
contributor for heavy elements. Case E is therefore slightly warmer than case
 A model and its BCZ is at a lower temperature. At $\sim 6$ Myrs case A BCZ is 
located at $0.3 M_{\odot}$ where $T=3.99\,10^6 K$ case E BCZ is located
at $0.4 M_{\odot}$ where $T=3.72\,10^6 K$. On the other hand 
by scaling every metal on oxygen we underestimate iron-peak
element fraction (and effects) in the atmosphere where they 
could provide up to 30 \% of free electrons. If the right iron-peak element fraction
was considered in the atmosphere, electron density would be enhanced. Convection 
would start closer to the surface and would therefore be less efficient. So we can 
expect that our values of BCZ temperature are in case E a bit too high. 
Figure 8 shows however that the changes among metal fractions are
mainly felt at BCZ. Difference in temperatures as a function of mass never exceed $5\%$
and changed opacities mainly affect BCZ but not the global structure.

\begin{figure}[ht]
\centering
\rotatebox{90}{\includegraphics[width=6cm]{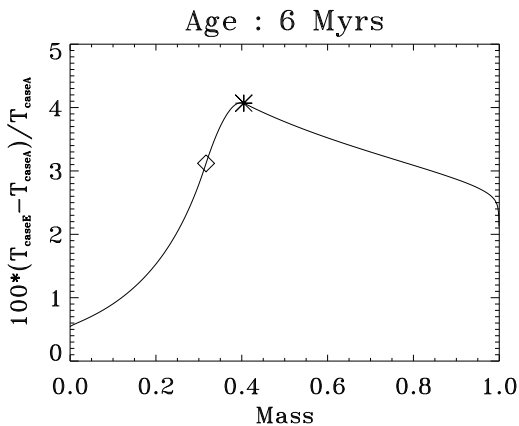}}
\caption{Relative temperature difference at a given depth between case E and case A models. Diamond and cross respectively locate BCZ of case A and E models.}
\end{figure}

\section{MACROSCOPIC EFFECTS: CONVECTION AND ACCRETION}

\subsection{Efficiency of convection}

The usual mixing-length theory parameter value relies on solar calibration.
There is no reason to believe it is universal. Recent work based on
hydrodynamical simulations (Ludwig et al. 1999) has investigated possible
calibrations of mixing-length for solar-type stars. This work shows a
dependence of $\alpha$ on effective temperature and surface gravity. For
temperatures between 7100 and 4300 K and surface gravities 2.54 $<$log g$<$ 4.74
$\alpha$ varies from 1.3 to more than 1.7. Using this work we find
negligible evolution of $\alpha$ on the Hayashi track for a 1 M$_{\odot}$
object where surface gravity varies significantly but effective temperature remains
almost constant. On the contrary once the radiative core becomes important
 and the star leaves the Hayashi track, the increase of the
effective temperature could justify a change of the  $\alpha$ value. For
a solar composition model this occurs at about 12 Myrs ie
slightly before the end of the $^7Li$ burning phase. In order to follow the
effective temperature impact on $\alpha$ we have computed models of 1 M$_{\odot}$
 stars for Pleiades, solar, and Hyades composition. Effective temperature
being sensitive to composition age and $\alpha$ parameters vary with
both of them. Temperature differences on the Hayashi track are
approximately 70 K if one changes from solar to Hyades composition.
In region of interest in temperature and gravity, differences in $\alpha$ are not
significant over such a narrow temperature range and the same $\alpha$ can be
used. After leaving the Hayashi track temperature differences increases up to 200 K
but the star evolves towards solar main sequence conditions in a zone where there
is a 'plateau' in $\alpha$ (Ludwig et al. 1999). Here again one can adopt a unique
value so that for present study the change in  $\alpha$ corresponds to
temperature and not composition. On this point Ludwig et al.(1999) agrees with
Fernandes et al. (1998) results  who found that $\alpha$ is almost
constant in the Sun and 4 low-mass stars systems having metallicities
between solar and -0.31 dex, and helium content from 0.25 to 0.28. For these
reasons we use the same values in the mixing length parameter whatever the
composition. Pleiades and solar composition being
close, we only made distinction between this group and Hyades.
Between these two groups composition determines the time spent on the Hayashi track. 
We have considered three different $\alpha$ values: on the Hayashi track, when the
radiative stratification begins to influence surface temperature and on 
the main sequence where the value results from usual solar calibration. Following
Ludwig et al. advices we have adopted $\alpha$ values they proposed after
reestimate by 0.1 and use their values as scaling factors. If the star
has solar or Pleiades composition, $\alpha$=1.935 before 15 Myrs ; $\alpha$=1.85  between 15
and 22 Myrs and $\alpha$=1.766 after 22 Myrs  which is our
solar-calibration main sequence value. If the star has Hyades composition, we
use the same values but at different times and change the limits to 20 Myrs and 28 Myrs.
Higher metallicity results in slower contraction and evolution.
In Hyades case only the initial value of $\alpha$ is important 
because after 20 Myrs $^7Li$ burning is over. 

Results of the hydrodynamical
calibration must be considered cautiously. Firstly, because low temperatures
molecular opacities are not included in Ludwig et al. calculations and we are presently
exploring low effective temperatures regions. Secondly a
recent result about $\iota$ Pegasi binary systems calibration predicts opposite
evolution of $\alpha$ with effective temperature (Morel et al 2000).
Thirdly 2D calculations have to be confirmed by 3D
hydrodynamical calculations.
\\
In these calculations, $^7Li$ depletion is increased (table 2, case B) as
convection zone extends deeper to higher temperatures with more efficient convection.
We note nevertheless that such modification leads to less lithium destruction than 
using a full spectrum of turbulence as Ventura et al. (1998).

\subsection{Mass accretion}

Observations of T Tauri stars accretion luminosities
lead to mass accretion rates spanning from a few $10^{-8}$
 up to a few $10^{-6} \, M_{\odot}$/yr (Hartigan 1995, Gullbring et al. 1998).
This wide range probably originate from real star to star
differences altough such low accretion rates are very difficult
to evaluate (Hartmann 1997) and suffer from large uncertainties.
Accretion can affect $^7Li$ stellar
 photosphere abundances in mainly three different
ways. First, it has a structural impact as it modifies
the stellar mass, consequently the gravitational potential then stratification
change. Secondly, providing ISM abundance material
to the surface of the star, has a direct chemical
impact. Thirdly, accretion should also change stellar boundary
conditions which is probably the most difficult part of
the accretion phenomenon to modelize. 

Our hydrostatic calculation
takes accretion into account in a crude fashion. The accreted mass modifies only 
the external layers of the star and does not directly affect global surface
boundary conditions (pressure, temperature, luminosity). Mass is simply
added on the outermost layer of the star.
We limit our study to accretion rates below $10^{-7} \, M_{\odot}$/yr
 and following Hartmann (1997), we consider global
accreted mass of a few $10^{-2} \, M_{\odot}$. 
As recent hints suggest that accretion could last longer than usually believed
 and perhaps deal with larger accreted masses (Muzerolle et al. 2000), so
we will then also investigate the effects of 0.1$M_{\odot}$
 mass accreted.

\subsection{A low global accretion mass}

We first evaluate $^7Li$ photosphere
variations through accretion by simply considering
different nonaccreting stellar masses. For a 2\% accretion
 mass, if the structural effect is dominant,
 final minimal $^7Li$ photosphere fraction
is the one of 0.98 $M_{\odot}$ non-accreting
star. This minimal fraction is 1.92, to compare with
2.13 for a 1 $M_{\odot}$ for a solar composition.
Then the maximal fraction can also be evaluated, if we
 consider that the chemical mixing is dominant.
This value is obtained in considering a 1$M_{\odot}$ star but artificially
increase $^7Li$ photosphere fraction by diluting
in external CZ the $^7Li$ mass contained in 0.02$M_{\odot}$ of
ISM material. The $^7Li$ maximal fraction is 2.61 for a 1 $M_{\odot}$.

A real accreting star has a lower mass all the way through
accretion phases and depletes both its initial lithium and
 lithium it receives from accretion.  
So, in the following we consider accretion impact on $^7Li$ burning
for solar composition star and 2$\%$ $M_{\odot}$ total accreted mass 
at varying temporal rates. Starting with a 0.98 $M_{\odot}$ stellar
 object we consider a 'fast accretion' rate model of $10^{-8}M_{\odot}/\rm yr$ during
2 Myrs, we get a lithium content of 2.12 dex and then a 'slow accretion' rate model
$10^{-8}M_{\odot}/\rm yr$ and $10^{-9}M_{\odot}/\rm yr$
during 1 and 10 Myrs respectively which gives a lithium content of 2.08 dex.
This kind of simulation does not affect depletion by more
than 0.05 dex. The general trend of accretion is to lower $^7Li$
fraction so we can conclude that structural effects are predominant 
over chemical effects. The effect is neither sufficient to 
 recover the agreement with observational
 $^7Li$ fraction nor to explain the spread in young clusters.

Accretion will have different consequences on lithium burning
if it could last long enough ie after major $^7Li$ pre-MS depletion phase.
Near IR excess of very low accretion
rates of $10^{-9} M_{\odot}/\rm yr$ are currently not dectected
(Hartigan, Edwards, \& Ghandour 1995).
The simple evaluation we make just above illustrates that a variation of only very
few percent of stellar mass could have a nonnegligible
chemical impact because after 10 Myrs external convection
zone is 10 \% or lower of the stellar mass. Then, an accretion rate as
low as $10^{-9} M_{\odot}/\rm yr$ could significantly change
ZAMS lithium surface fraction.

\subsection{Larger accretion rates applied to the Pleiades composition}

 $90 \%$ or more of stellar final mass is accreted during short 
(less than one myr) class 0 and class I stages who
 extend up to a few $10^5$ years before classical T Tauri
 phase (Andre, Ward-Thomson, \& Barsony 1999 and references therein).
We consider here both different final masses from 0.9 to
1.1 $M_{\odot}$ and different globally accreted masses ie $2\,10^{-2}$
or $10^{-1}$$M_{\odot}$. The stars now have Pleiades composition. 
As we have seen in the solar case, structural effect
of accretion is maximum when accretion is fast, we therefore restrict
our computation to 'fast' accretion process: $10^{-8}$ or
 $5\,10^{-8} M_{\odot}.\rm yrs^{-1}$
during the first Myr and then $10^{-9}$ or $5\,10^{-9} M_{\odot}. \rm yrs^{-1}$ from
1 to 11 Myrs. In these conditions, accretion still increases $^7Li$
depletion and confirms solar trend. In table 4 we
give lithium abundance relative to hydrogen
as function of mass and with or without accretion in case of Pleiades. The lower the mass
the higher the accretion impact. Accretion could therefore
explain initial lithium dispersion that also seem to increase
towards low effective temperatures.

\begin{table}[ht]
  \begin{center}
    \caption{Lithium surface abundances at 70 Myrs for accreting and
         non-accreting Pleiades composition models.}\vspace{1em}
    \renewcommand{\arraystretch}{1.}
    \begin{tabular}[h]{lccc}
        Mass & 1.1$M_{\odot}$ & 1$M_{\odot}$ & 0.9$M_{\odot}$ \\
      \hline
        Effective\\
        temperature & 6050 & 5720 & 5350\\
      \hline
        Observed mean\\
        lithium abundance & 2.92 & 2.70 & 2.57\\
      \hline
        No accretion\\
        models & 2.95 & 2.58 & 1.66\\
      \hline
        Global accreted\\
        mass = 0.02$M_{\odot}$ & 2.94 & 2.56 & 1.59\\
      \hline
        Global accreted\\
        mass = 0.1$M_{\odot}$ & 2.86 & 2.39 & 1.24\\
      \hline
      \end{tabular}
  \end{center}
\end{table}

\begin{table}[ht]
  \begin{center}
    \caption{Evolution in $M_{\odot}/yr$  of required accretion rates to reach lithium surface abundances at Pleiades age.}\vspace{1em}
    \renewcommand{\arraystretch}{1.}
    \begin{tabular}[h]{lcc}
        Age (myrs) & Accretion 1$M_{\odot}$ & Accretion 0.9$M_{\odot}$ \\
     \hline
        2 & $6.3\,10^{-8}$ & $7.9\,10^{-8}$\\
        3 & $6.3\,10^{-8}$ & $7.9\,10^{-8}$\\
        4 & $4.0\,10^{-8}$ & $6.3\,10^{-8}$\\
        5 & $3.1\,10^{-8}$ & $4.0\,10^{-8}$\\
        6 & $1.6\,10^{-8}$ & $2.0\,10^{-8}$\\
        7 & $10^{-8}$ & $1.2\,10^{-8}$\\
        8 & $6.3\,10^{-9}$ & $7.9\,10^{-9}$\\
        9 & $4.0\,10^{-9}$ &  $4.0\,10^{-9}$\\
        10 & $2.5\,10^{-9}$ & $2.5\,10^{-9}$\\
        15 & $1.6\,10^{-10}$ & $2.5\,10^{-10}$\\
        20 & $5\,10^{-12}$ & $2.5\,10^{-11}$\\
      \hline

      \end{tabular}
  \end{center}
\end{table}

For a comparison, we estimate observational lithium abundance  dispersion in
the data set of Soderblom et al. (1993)  we divide into 200 K bins.
 For each bin we then compute mean abundance value and dispersion. Between
 4200 and 6000 K, we find that dispersion varies from 0.65 to $\sim 0.1$ dex
exhibiting the well-known general trend to decrease when tempature increases.
Around T$_{eff}$= 5720 K (1 $M_{\odot}$) it is $\sim 0.12$ while around T$_{eff}$= 5350 K
 (0.9 $M_{\odot}$) it is $\sim 0.32$. As can be seen in table 4 these results are
qualitatively and quantitatively comparable to predicted dispersion resulting from
accretion of 10 \% of M$_{\odot}$. We conclude that
accretion could explain early main sequence dispersion. Results suggest also
that the more a star accretes the more
it depletes $^7Li$ because once again in these cases structural effects dominate chemical effects.
$^7Li$ is not refreshed at a sufficient level to compensate for the additive burning
due to the lower mass.

We now evaluate the minimal accretion rate necessary
to counteract mass effect for stellar masses of 0.9 and 1 $M_{\odot}$ (table 5).
Every million years photosphere $^7Li$ fraction disminishes by a given amount and knowing
convection zone mass and ISM $^7Li$ fraction we compute a 'nominal' mass accretion
rate necessary to exactly compensate losses. One could however argue that 1 $M_{\odot}$
 Pleiades have not kept initial $^7Li$ fractions.
They indeed depleted from ISM value $\sim 3.2$ down to $\sim 2.8$. Taking this
into account we compute a new accretion rate in the following way : if a fraction $\alpha$ of
'nominal' accretion is provided every Myr only $1-\alpha$ of the quantity
of depleted $^7Li$ with no accretion will finally be depleted. We find
that new accretion rates should be lowered from 'nominal' ones by 0.32 dex and we provide
results in table 5. Results do not
really change within considered mass range. When compared to observations of present star
forming regions (Calvet \& Gullbring 1998; Muzerolle et al. 2000)
these accretion rates seem slightly too high. The required accretion rates
exceed strongest reported accretions observations by approximately 0.5 dex
which brings us close to the observational upper limits.

It is clear that conclusion on the accretion rates requires improved detection
of mass accretion estimates to rates as low as $10^{-9} M_{\odot}\rm/yr$
for a large number of very young clusters, especially for low mass stars 
and corresponding photospheric lithium content.
 Therefore preceding results are mainly
illustrative rather than conclusive. Moreover, the accretion process has to be
approached hydrodynamically and maybe in considering periodic
phenomena, to take into account the modified boundary conditions.

\section{MACROSCOPIC EFFECTS: THE INFLUENCE OF THE ROTATION}

Regarding solar-like stars, open clusters $^7Li$ abundances
 suggest a general depletion over main sequence
 which is in contradiction with standard evolution codes
results. We will examine in this section the role of the internal rotation
firstly on the pre-mainsequence structure and secondly on the possible tachocline
mixing process inside radiation zone. 

\subsection{Rotation structural effects}

Within the last decade many theoretical works have been led in the
 field of angular momentum transport in stellar interiors.
Such works encounter difficulties to explain rotation velocity evolution.
 For instance models invoking hydrodynamic angular momentum transport 
 (Pinsonneault et al. 1989) predict strong differential rotation in solar interior
 which is contradictory with helioseismology results. Models with
angular momentum transport induced by internal 
magnetic fields (Keppens, Mac\,Gregor, \& Charbonneau 1995)
 can fit observations with a core-enveloppe coupling time of the order of
 10 Myrs. They however encounter problems in reproducing the slowest rotators in open
clusters and early MS rotational evolution. 

Core rotation could still significantly differ from surface rotation.
Rotation rate is presently observed to be constant in most of the solar radiation
 zone and varies with latitude in solar convection zone. However it could
 have been a varying quantity in the initial stellar radiative and/or convective
 interior. We investigate here purely structural effects
of varying internal rotation and examine the consequences 
on effective temperature, luminosity, and lithium content.
The models we consider have Hyades composition where the
stars below 1$M_{\odot}$ all have experienced an initial rotational history
 increase until ZAMS followed by a strong decrease (see section 5.2). On the
other hand they are still at the beginning of ZAMS so there is probably no long term MS
effects on lithium. This cluster therefore seems
to us quite appropriate to study early rotational effects. We compute three
models in the case of very short-lived circumstellar disk of 0.5 Myrs. 
The first model (SBR), assumes solid body rotation throughout the star.
The second (DR), assumes complete decoupling between radiative and convective zones
and the last model (FR) assumes solid body rotation in convection zone and
 every stellar mass shell keeps its initial angular momentum in radiative
 stratification part. If the first model corresponds to zero coupling time
between convection and radiation zones, this coupling time is infinite in 
the latter ones.

A well known structural effect of rotation is the decrease
of the effective temperature and to a less extent of stellar
luminosity (Sills, Pinsonneault, \& Terndrup 2000). For the SBR stellar model
 we compute equatorial velocity of 74 km/s on ZAMS ($\sim 57 Myrs$).
This gives rise to a temperature decrease of only 23 K in comparison with the
 nonrotating model. Under analog conditions, the  
 polynomial formula of Sills et al. gives a discrepancy of 15 K. The same
formula at maximal speed age (110 km/s for $\sim 30 Myrs$)
leads to 50 K where we find 65 K in difference.
Differences never exceed 0.03 dex in luminosity
 and 2 \% in radius at given time. The expected consequence
of such a behavior over convection is a larger extension and 
higher BCZ temperature and density responsible of the higher depletion
rate in $^7Li$. Temperature at the BCZ increases strongly when the
effective temperature lowers. This mass-lowering is effective when the star reaches
its maximal rotation rate around 30 Myrs. Yet before $\sim 7 Myrs$
the situation is reversed and rotation models surprisingly exhibit
lower BCZ temperature and density. A maximum difference in temperature of $2\,10^4$K
appears at 3 Myrs. It seems indeed
 that early rotational effects over BCZ are opposite to
 what they become afterwards. This trend changes the 
 temperature sensitivity on $^7Li$ depletion rate.

\begin{table*}[ht]
\caption{ Impact of internal rotation model on lithium fraction in dex for 
solar mass and Hyades composition star }\vspace{1em}
\renewcommand{\arraystretch}{1.}
\begin{tabular}{lllccc}
\hline
$\tau_{disk} in Myrs$   &No rotation    &SBR     &DR     &FR\\
 & & & &\\
Lithium at 7 Myrs               & 2.06  & +0.11  & +0.11  & -0.01\\
\hline
Lithium at 20 Myrs              & 0.96  & +0.07  & +0.08  & -0.07\\
\hline
Lithium at 0.7 Gyrs             & 0.91  & +0.07  & +0.07  & -0.05\\
\hline
\end{tabular}
\end{table*}

Quantitative variation of lithium fraction between
slowly ($\tau_{disk}$=3 Myrs) and rapidly ($\tau_{disk}$
=0.5 Myrs) rotating models is very small : 0.1 dex (Table 6).
No difference exists between SBR and DR models because
the rotation is the same before 20 Myrs. In fact, the rotation rate,
anywhere in the star, is independent of 
any coupling time at least until 20 Myrs and at this age $^7Li$ pre-MS 
depletion is over. There are two reasons for that. 
Firstly radiation zone stratification is still 
hardly under-adiabatic until 30 Myrs as we already explained in section 2.1.
Secondly stellar wind loss represents only $12\%$ of initial
momentum at 20 Myrs (vs $\sim 99\%$ at solar age). Contraction 
of the whole star therefore stays homologuous and coupling-time
 between zones has no impact. Unless initial conditions are 
different from solid rotation for the initial fully convective body there
are no requirements to bother about angular momentum loss or exchanges
 during early pre-MS $^7Li$ burning phases.

In the preceding lines we have been concerned by structural effects
as far as they result from a correction to local gravity. We do not
note any significant impact of these effects on the present problem.
Now there are others means by which rotation might affect stellar
structure. Convective movements are sensitive to rotation through 
coriolis force. This effect is known for long in earthly atmospheric 
global circulation notably in the intertropical region. Siess \& Livio 1997
 suggest that convective cells could be twisted by rotation allowing 
therefore a smaller $\alpha_{mlt}$ to the convective zone. As long
as precise hydrodynamical computation are led on that phenomenon
 it is however difficult to say much more about this solution.

\subsection{The turbulence at the base of the convection zone}

We consider now that the rotation induces an hydrodynamical
 instability in the tachocline layer at the top of the 
radiation zone (Spiegel \& Zahn 1992). Such an instability has been studied by 
BTCZ99 to interpret the helioseismic results (Kosovichev et al. 1997)
and could partly be at the origin of lithium destruction during the mainsequence.
The introduction of a time dependent turbulent term in the 
diffusive equation permits a better agreement on solar 
photospheric light elements between observations and models.
BTCZ99 remark that in the present Sun, tachocline
 mixing is thin enough not to deplete $^9Be$.
 However tachocline mixing depends on rotation and differential 
rotation which is poorly known in pre-MS stars.
\\
We reexamine this process introducing a more realistic rotation history in pre-MS.
A 1$M_{\odot}$ model of Hyades composition and including tachocline
mixing will not deplete more than 0.03 dex in $^9Be$ from formation
 until ZAMS and none afterwards. Beryllium depletion does not occur 
although rapid rotation and high BCZ temperatures. It is the reason 
why we do not discuss $^9Be$ in this study.

Skumanich law (Skumanich 1972) has been used to 
infer rotational time evolution in mainsequence (as in BTCZ99).
But such a law cannot be extend to young pre-MS stars
which initially rotate slowly and then experience strong
acceleration toward ZAMS.
We reevaluate here tachocline mixing in using a more 
realistic rotation law that should apply over both pre-MS
 and MS. So we adopt the approach of Bouvier, Forestini,
 \& Allain (1997, hereafter BFA97) to calculate different sets of rotation
evolution for 1 $M_{\odot}$ stars under the three
 following assumptions : (i) stars rotate as solid bodies,
 (ii) stars are locked to a defined angular velocity  
as long as they interact with initial surrounding
 disk, (iii) magnetic wind braking acts all along pre-MS and MS and
depending on rotation speed produces varying angular
 momentum loss rate:

\begin{equation}\label{eqn: coef}
(\frac{dJ}{dt})_w=-K{\Omega}^3({\frac{R}{R_o}})^{1/2}({\frac{M}{M_o}})^{-1/2}\,\rm if \, \Omega
< {\omega}_{sat}
\end{equation}

\begin{equation}\label{eqn: coef}
(\frac{dJ}{dt})_w=-K{\Omega}{{\omega}_{sat}}^2({\frac{R}{R_o}})^{1/2}({\frac{M}{M_o}})^{-1/
2}\,\rm if \, \Omega > {\omega}_{sat}
\end{equation}

Figure 9 presents corresponding equatorial velocity
 evolution for solar mass and composition.
\begin{figure}[ht]
\centering
\rotatebox{90}{\includegraphics[width=6cm]{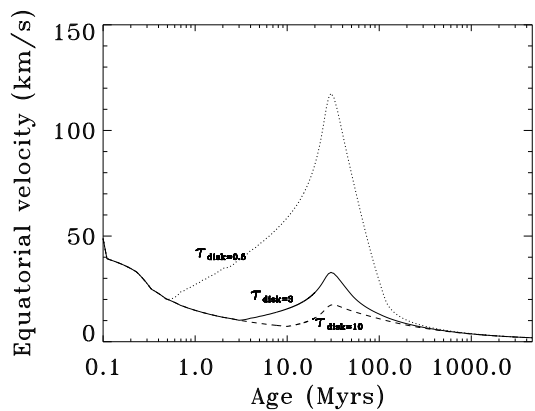}}
\caption{
Equatorial velovity vs age for stars of solar mass and composition. From fastest to
slowest rotators $\tau_{disk}$ is 0.5,3,10 Myrs.
 }
\end{figure}
$\omega_{sat}$ and K are adjusted to fit observations
 of surface rotation rates.
 Surface magnetic field in solar type stars seems to increase
with $\Omega$ up to $\sim 10 \Omega_{o}$ and then saturates
 (Saar 1996). Saturation value determines a transition in
the efficiency of the wind to brake the star which has consequences over
early MS rotation rates. Following BFA97 we adjust
 $\omega_{sat}$=14 $\Omega_{o}$. 
 K determines the rotation rate at solar age.
Recent helioseismic measurements (Corbard et al. 1997) give
internal solar rotation rates. A latitudinal
 dependence of rotation frequency is observed in convection
zone, from 460 nHz at equator down  to 370 nHz
at 60 degrees latitude. Radiative zone then experiences rigid
rotation at the 430 nHz level and at least as deep as 0.4
 solar radius. In present solid rotation models, we adjust
 the rotational evolution laws so that the frequency is 430 nHz
at solar age. To reach this rotation rate at actual solar age
 we adjust K to 3.25 $10^{47} g.cm^2.s$ in the expression
 of the braking law. This value is a little larger than the one used by BFA97
 (2.7 $10^{47} g.cm^2.s$) as we tend towards a slower
 rotation velocity than surface equatorial :
$2.7\,10^{-6} rd.s^{-1} $ instead of $2.9\,10^{-6} rd.s^{-1}$.

We then consider three different $\tau_{disk}$ durations
 for star/disk coupling time : 0.5, 3, and 10 Myrs. 0.5 Myrs corresponds to a star
 that ceases disk locking evolution on its birthline
or a few 0.1 Myrs after depending on mass and composition.
 Three Myrs is the median disk lifetime as estimated by BFA97. 
Finally 10 Myrs corresponds to a persistent disk.
At this age only 10 to 30 $\%$ of young stars still show
the IR and mm radio emission expected if an optically thick
disk is present (Strom 1995). The star is kept on a velocity
 of $9.1\,10^{-6} rd.s^{-1}$ until it uncouples from disk.
Then owing to gravitational contraction it accelerates up
to $ 2.70\,10^{-5} rd.s^{-1}$ and  $1.75\,10^{-4} rd.s^{-1}$
 after $\sim 30$ Myrs for $\tau_{disk}$ equal 10 and 0.5 Myrs respectively.
Such rotation rates correspond to equatorial velocities that
 span from 20 km/s to 120 km/s. Afterwards the stars decelerate
rapidly towards the same low velocity whatever disk lifetime. At the
Hyades age the stellar equatorial velocity is in the narrow
 range from 4 to 6 km/s.

Metallicity has direct impact on rotation evolution.
We evaluate rotation history for three sets
 of metallicities : Pleiades,
 Sun, and Hyades. From the former to the later composition contraction
 time towards ZAMS increases. Peak velocities are reached later and are slightly
lowered mainly due to higher moment of inertia and radius. A 1$M_{\odot}$ pleiad reaches
maximal velocity at 29 Myrs whereas its Hyades conterpart reaches it at 33 Myrs. 
Moreover radius
being a bit larger the wind braking affects more surface
rotation. For a typical disk lifetime of 3 Myrs we estimate peak velocity
to be 120 km/s and 113 km/s for Pleiades and Hyades compositions.
These values are reached around 29 and 33 Myrs respectively.

Purely structural effects of rotation have been included the same
way as in the preceeding section.
Rotational centrifugal acceleration effects are added to gravity.
This modifies the equation of hydrostatic equilibrium and
radiative gradient. At radius r and angular velocity $\omega$,
 the present stellar code integrates the 'mean'
effect of rotation by substracting $2r\omega^2/3 $
to gravitation. Then rotation velocities are also taken into account in a
 macroscopic rotationally induced diffusion coefficient $D_{T}$ (equation 14 in
BTCZ99) and
tachocline thickness d (equation 11 in
BTCZ99) that determines tachocline mixing efficiency.
These coefficients are assumed to follow the scaling laws 
(suggested by BTCZ99):

\begin{equation}\label{eqn: coef}
D_{T} \, \alpha \, \Omega^{0.75\pm0.25}
\end{equation}
\begin{equation}\label{eqn: coef}
d \, \alpha \, \Omega^{(1.3\pm0.1)/4}
\end{equation}

\subsubsection{The solar case}

Rotation induces a strongly increased depletion of photospheric $^7Li$. 
 Structural modifications do
not play any significative role on pre-main sequence as well as on
main sequence unless the stellar disk lifetime is very short. For a
3 Myrs disk lifetime we note no differences exceeding 0.1 dex in $^7Li$
between a model that includes rotation structural changes and a model
that does not. Stellar structures and evolutions are similar. On the contrary
rotationally induced mixing dramatically changes $^7Li$ history.
The 3 Myrs disk lifetime star experiences an
enhanced destruction of $^7Li$ during radiative
core developpement phase when compared to non rotation model.
$^7Li$ abundance is lowered from $\sim 2.1$ dex without rotation down to
$\sim 1.8$ dex. This increases $^7Li$ depletion during pre-main sequence
in comparison with previous calculations using only the Skumanich law (BTCZ99). 
But the final result is not much altered.
 $^7Li$ depletion depends on disk lifetime $\tau_{disk}$. Variation
is important between $\tau_{disk}$=0.5 and 3 Myrs but more moderate
for longer disk lifetimes (table 7, figure 10, and 11).

\begin{table*}[ht]
\caption{Lithium fraction in dex for solar mass and composition star as a function of
 circumstellar disk lifetime.}\vspace{1em}
\renewcommand{\arraystretch}{1.}
\begin{tabular}{lllccc}\hline
$\tau_{disk} in Myrs$                           &0.5            &3              &10
&20\\
\hline
Lithium at ZAMS                 & 1.34  & 1.80  & 1.90  & 1.92\\
\hline
Lithium at 0.7 Gyrs             & 0.84  & 1.43  & 1.56  & 1.61\\
\hline
Lithium at 4.6 Gyrs             & 0.35  & 0.95  & 1.10  & 1.16\\
\hline
\end{tabular}
\end{table*}

Typical scatter between long and short lived disk is 0.5 dex and
tends to grow slightly on main sequence. This does not agree
with open-cluster observations that shows no dispersion in
$^7Li$ abundances from Pleiades to Hyades as can be seen
on previous plots. On the other hand spread reappears
in much older clusters as M67 (Jones, Fischer, \& Soderblom 1999). With a
photospheric lithium of 1.1 dex, the Sun is expected 
to have experienced a long-lived disk stars (figure 10), 
if this approach is correct. However one has to be very 
extremely cautious when drawing such conclusion. Pre-MS 
depletion phase is overestimated as young open-clusters observations
suggest.

\subsubsection{The Hyades case}

Tachocline thickness and macroscopic diffusion coefficient
depends on the rotation history and the metallicity.
However if scaling laws of equations 3 and 4 are correct, we
compute that differences in metallicity of 0.1 dex do
 not induce large enough modifications in rotation
speed to significantly change tachocline diffusion
coefficient or thickness in pre-MS.
The general behaviour noticed for the Sun
is also observed for this cluster. 
But the computation shows that higher velocity rotating stars 
do exhibit larger $^7Li$ depletion,
this does not agree with present observations, indicating that
rapid rotators have higher $^7Li$ rates as their slower
counterparts (Soderblom et al. 1993 ; Garcia Lopez, Rebolo, \& Martin 1994).
This point suggests that tachocline mixing, as it is now introduced
in calculations, and probably any mechanism that would mix
radiatively stabilized layers on pre-MS, could be partly inhibited.
It is noteworthy that low mass star rotation are presumably evolving slower. 
A 0.8 $M_{\odot}$ star should reach peak velocity at 50 Myrs. 
Moreover such a star will retain its angular momentum 
longer so that tachocline mixing retains longer efficiency. This brings the
star the wrong way for it is the lighter stars that present the largest discrepancy in
$^7Li$ content between current models and observations.

\begin{figure}[ht]
\centering
\rotatebox{90}{\includegraphics[width=6cm]{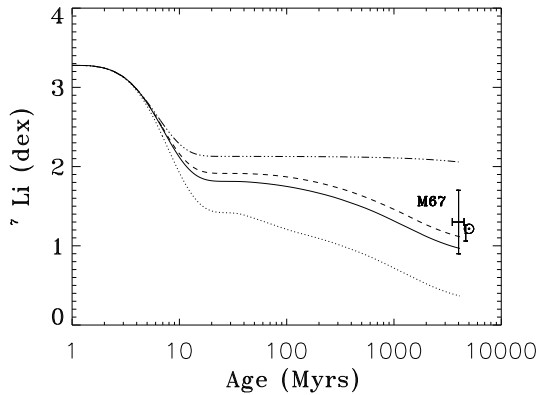}}
\caption{ $^7Li$ surface abundances in solar mass and composition
stars. From the lowest to the highest depletion rate : first model
only includes microscopic diffusion and three following models add tachocline
mixing with disk coupling time 10, 3, and 0.5 Myrs respectively. Also plotted
M67 and solar surface lithium abundances. Every model exhibits the same effective 
temperature of 5777 K at 4.6 Gyrs (table 2 : solar models).}
\end{figure}

\begin{figure}[ht]
\centering
\rotatebox{90}{\includegraphics[width=6cm]{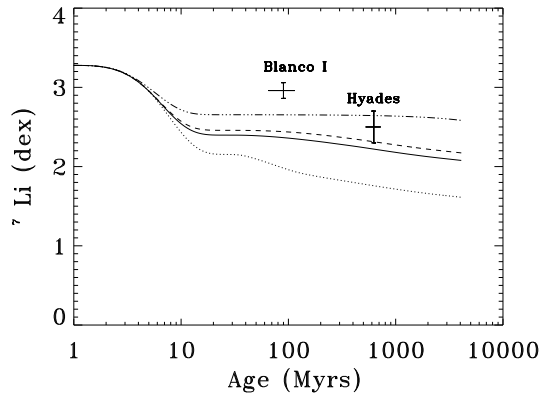}}
\caption{ $^7Li$ surface abundances in solar mass and Hyades case E composition
stars. Same presentation than above. Also plotted
Blanco I and Hyades surface lithium abundances. Microscopic diffusion model
has an effective temperature of 5802 K at 625 myrs. At this age all other models
exhibit very similar effective temperatures (from 5800 to 5802 K)  
}
\end{figure}

\section{SUMMARY AND PERSPECTIVES}

This study confirms that lithium offers an extremely interesting 
insight over solar-like stellar structure and evolution due to its low depletion 
temperature.
There is presently a discrepancy between computations of classical
stellar models and observations. Classical models predict no lithium depletion
on main sequence whereas it seems to be observed in open-cluster lifetimes.
It is reasonable to think that lithium evolution on main sequence 
is connected to some slow mixing rotationally induced process(es) that 
occurs at 
the top of the radiative zone. In the specific case of our Sun such a process is 
supported
by differences between theoretical and measured sound speed in this
part of the star (BTCZ 1999) and by the agreement between photospheric 
observations and predictions for helium, metals, and lithium. 
During pre-mainsequence, the lithium problem
is reversed and classical uptodate theoretical models generally predict too strong
 depletion. We confirm that this depletion strongly depends on metallicity,
 in the calculation. It is not evident that it is
currently observed in young open-clusters. In this 
regard Coma Ber and Blanco I rather suggest an age dependence. 
Moreover, early dispersion in lithium abundances has to be explained.

In this paper, we have studied the impact of different microscopic or 
macroscopic processes on pre-MS lithium depletion. First we note the important 
role of the choice of meshes and time steps 
in the very early phase (before 10 Myrs). Secondly, we remark  that
the interpretation of data is presently difficult as $^7Li$ depletion 
in open-clusters is a strong function of effective temperature and it is
therefore very difficult to reliably relate an observed lithium fraction 
to a given stellar mass just because this effective temperature depends  
on detailed metal fractions that still have to be accurately determined 
(see figure 6 compared to figures 7 and 11).

Microscopic processes are mainly dominated by opacity coefficients
 and therefore through metal and helium fractions.
We experience in models that the level of 
sensitivity to composition (especially on oxygen and iron but not only) 
proves how difficult the present situation is. This 
suggests that
the dispersion in lithium abundances could be related to small dispersion 
in metal fractions or among metal fractions. This is all 
the more true as part of the observed scatter is surely related to 
observational questions. Rotation history might 
induce such difference but on the other hand King, Krishnamuthi
, \& Pinsonneault 2000 note the $^7Li$ abundance determination to be correlated to 
(and probably biased by) chromospheric activity. Genuine dispersion
could also be related to larger helium fraction variations which
can not be ruled out even over small scales in the ISM
 (Wilson \& Rood 94 and references therein). In the solar case,
 where all the abundances are known within 10\%, we get the good order of 
magnitude
 of lithium depletion in adding a macroscopic effect to the implicit microscopic 
 effect in pre-mainsequence. We do not observe also a large problem in Hyades
 when using observed oxygen and iron are measured. Nevertheless there is a contradiction
between the large dependence of lithium depletion in theoretical models in comparison
to the apparent quasi metallicity independence of the abundances in young open-clusters.
To solve this problem we encourage
 measurements of photospheric abundances as carbon, oxygen, silicium, and 
iron on young T-Tauri stars.
Similarly helium fraction needs to be carefully checked within 
clusters and if possible from star to star. Such additional data
 can be deduced from binary systems, high mass stars, and importantly
 in solar-like stars thanks to the future asteroseismology developments.

Macroscopic processes are at play through accretion, rotation, 
and convection modelling. Such processes are now crucial to 
understand accurately stellar evolution (Pinsonneault 1997).
 Accretion is likely to explain
dispersion in $^7Li$ fraction measurements, but it fails
to bring mean abundances to an acceptable level. In Pleiades case,
required accretion rates at a given time are at 
the upper limit of observations for young T Tauri.
To improve significantly the situation they should last 
so long that total accreted mass would indeed be $\sim 20-30 \%$ of 
solar mass during the first Myrs following star formation. 
On the other hand it is also possible for young stars not to
accrete from classical gas \& dust disks but from planetesimals of protoplanetary
systems which are not detectable today. Such material infall
could supply lithium and others metals in low mass convective enveloppes of post-TTauri
stars.    
 
Rotationally induced mixing gives satisfactory results regarding
photospheric solar lithium abundance. These results are moreover
consistent with Hyades open-cluster observation when oxygen and iron fractions
are correctly ie separatedly considered.  
Rotation mixing process we consider is however unable to 
explain initial spread in $^7Li$ surface abundances. 
Treating more properly the fruitful MS tachocline mixing (BTCZ 1999)
in pre-MS leads to results not supported by observations 
(King, Krishnamurthi, \& Pinsonneault 2000 and references
 therein) : we find a rotation-$^7Li$ depletion correlation on ZAMS.
However it is important to keep in mind two points.
Firstly we recall that the present rotationally induced mixing is not related to
angular momentum transport in the stars. It therefore offers 
a possible but certainly partial vision of lithium MS evolution. ZAMS rotational velocity
 should affect $^7Li$ depletion because of subsequent angular momentum redistribution 
(Pinsonneault et al. 1999 and references therein). Secondly the rotationally induced 
mixing process
 we invoke has necessary a transitory regime that we did not take
into account. For these two reasons $^7Li$ depletion predictions 
of the tachocline mixing are more reliable between the Hyades and older clusters than absolute
predictions from star formation. Hyades solar-like stars are already slow rotators
and might have lost most of their angular momenta. From Hyades to older
clusters, lithium should traduce long term (stationary) mixing effects, and indeed
the tachocline mixing give rise to the observed characteristic time ($\sim 1 Gyr$)
 for $^7Li$ depletion on MS.

Finally it is possible that the convection parameter $\alpha_{mlt}$ is sensitive
to rotation and hydrodynamical simulations suggest it
should be changed along Hayashi track. The effect we mention is not very large, 
contrary to some previous studies but more work is needed to be sure. We have 
emphasized the absence of a well established radiative 
stratification
in central parts during phases preceeding 20 Myrs which is a tricky
point. The first phase of lithium burning occurs at the frontier 
between cooler convective medium and hotter radiative (but almost convective)
 medium that lies beneath. This means that very slight additional perturbation
 (through overshooting or instability for instance) could give rise to a much 
stronger lithium depletion. This also means that very slight 
stabilisation phenomenon, related to magnetic field as it 
is suggested by Ventura et al. (1998) could save a large fraction of lithium.
 
 All these remarks on macroscopic phenomena encourage complementary studies 
implying  hydrodynamics approach. This is also true to check if
the star is initially fully convective
and rotates as a solid body. The answers to those
questions can only be extracted from earlier 
phases that include protostellar collapse and 
take into account hydrodynamical processes such as accretion
in a consistent and more rigourous manner.

\begin{acknowledgements}
Acknowledments

Authors are grateful to Forrest Rogers and Carlos Iglesias from LLNL
 who made possible computation of opacity tables corresponding 
to non-solar repartition. They are also grateful to Jean-Paul Zahn
for enlightening discussion on stellar 
structure and hydrodynamic instabilities and to Sacha 
Brun for strong interaction. We thank Sylvain 
Turcotte for his help on opacity and ionisation states calculations 
and we also thank Jean-Pierre Chi\`eze Philippe Andr\'e and 
Jean-Paul Meyer for giving some helpful advises on the fields of young 
stars physics and cosmic chemical abundances. Finally we are 
grateful to the referee for very interesting and helpful advices
on the manuscript.

\end{acknowledgements}

\end{document}